\documentclass[12pt,preprint]{aastex}

\slugcomment{ Preprint: {\it Astrophysical Journal}}

\shorttitle{Energy Cascades in Astrophysical Plasma}
\shortauthors{Dastgeer \& Zank}

\newcommand{\be}{\begin{equation}}
\newcommand{\ee}{\end{equation}}

\newcommand{\eqs}[2]{Eqs. (\ref{#1}) \& (\ref{#2})} 
 
\newcommand{\Eq}[1]{Equation (\ref{#1})} 
\newcommand{\eq}[1]{Eq. (\ref{#1})} 
\newcommand{\fig}[2]{Figs  (\ref{#1}) \& (\ref{#2})} 
\newcommand{\Fig}[1]{Fig. (\ref{#1})} 
 \newcommand{\eqa}{\begin{eqnarray}}
\newcommand{\eeq}{\end{eqnarray}}  
\newcommand{\eqsto}[2]{Eqs. (\ref{#1}) to (\ref{#2})} 

\newcommand{\sig}{\sigma_{\rm ex}}
\begin{document}


\title{Energy Cascades in a Partially Ionized Astrophysical Plasma}


\author{Dastgeer Shaikh\footnote{Electronic mail : {\tt dastgeer@ucr.edu}} 
  \and 
Gary P. Zank\footnote{Electronic mail : {\tt zank@ucr.edu }}}
\affil{Institute of Geophysics and Planetary Physics (IGPP),\\
University of California, Riverside, CA 92521. USA.}



\begin{abstract}
A local turbulence model is developed to study energy cascades in the
interstellar medium (ISM) based on self-consistent two-dimensional
fluid simulations. The model describes a partially ionized
magnetofluid interstellar medium (ISM) that couples a neutral hydrogen
fluid with a plasma primarily through charge exchange interactions.
Charge exchange interactions are ubiquitous in warm ISM plasma and the
strength of the interaction depends largely on the relative speed
between the plasma and the neutral fluid. Unlike small length-scale
linear collisional dissipation in a single fluid, charge exchange
processes introduce channels that can be effective on a variety of
length-scales that depend on the neutral and plasma densities,
temperature, relative velocities, charge exchange cross section and
the characteristic length scales. We find, from scaling arguments and
nonlinear coupled fluid simulations, that charge exchange interactions
modify spectral transfer associated with large scale energy containing
eddies. Consequently, the warm ISM turbulent cascade rate prolongs
spectral transfer amongst inertial range turbulent modes. Turbulent
spectra associated with the neutral and plasma ISM fluids are
therefore steeper than those predicted by Kolmogorov's phenomenology.

\end{abstract}




\section{Introduction}

The local interstellar medium (LISM) surrounding the heliosphere is
warm ($\sim 7000^o K$) and consequently partially ionized. Although
the ionization fraction of the LISM is not conclusively established
\citep{slavin2006}[see also \citep{muller2006,zank2006}], it is
possible that the plasma number density is as small as $0.05 cm^{-3}$
and the neutral atomic H number density as much as $0.16
cm^{-3}$. Indeed, much of the interstellar medium is warm and
partially ionized. In the LISM, the low density plasma and neutral H
gas are coupled primarily through the process of charge exchange. On
sufficiently large temporal and spatial scales, a partially ionized
plasma is typically regarded as equilibrated; this is the case for the
LISM. However, the region between the bow shock of our heliosphere and
the heliopause \citep{zank1999} is not equilibrated because the charge
exchange mean free path (mfp) and the size of the region are
comparable. The neutral H distribution in the region is complex
\citep{pauls1995,zank1996,zank1996b,jacob2006} possessing an
approximately Maxwellian core that is broadened by hot H atoms produced
in the inner heliosheath \citep{zank1996b} and fast neutral H produced
in the supersonic solar wind \citep{zank1996b,zank1999}. Similarly,
the atmospheres of many stars, especially cooler stars, that are
embedded in a partially ionized interstellar medium will have outer
astrosheaths that are partially ionized. Several examples of such
partially ionized astropsheres have been observed using Lyman-$\alpha$
absorption measurements \citep{gayley1996, wood2000}.

Turbulent fluctuations in a partially ionized LISM are not only
potentially important in the context of the global heliospheric ISM
interaction, but are also instrumental to our understanding of many
astrophysical phenomena including the energization and transport of
cosmic rays, gamma-ray bursts, ISM density spectra etc. Besides the
ISM, a partially ionized plasma environment represent an important
component of the Saturnion magnetosphere, especially that part
influenced by mass loss from the moon Encephalus \citep{ip1997}, in
Neptune's magnetosphere e.g., \citep{hill1990}, and also in the
astrosphere of stars embedded in a cloud of neutral H or even some
stellar winds. For magnetosphere, the partially ionized plasma
sometimes comprises a mixture of water group and O neutrals and pickup
ions and solar wind plasma.  The partially ionized plasma environment
is also relevant to the study of neutron stars \citep{Potekhin2005}
where the plasma and strong magnetic field interaction still remains
an unresolved issue.

 The interaction of a neutral gas and plasma can be dated back to the
 seminal work by Kulsrud \& Pearce (1969) in the context of cosmic ray
 propagation, where it was shown that neutral component damps Alfv\'en
 waves. Neutrals interacting with plasma via a relative drag process
 results in ambipolar diffusion. Ambipolar diffusion plays a crucial
 role in the dynamical evolution of the near solar atmosphere,
 interstellar medium, and molecular clouds and star formation. For
 instance, Oishi \& Mac Low (2006) investigated the inability of
 ambipolar diffusion to set a characteristic mass scale in molecular
 clouds and find that substantial structure persists below the
 ambipolar diffusion scale because of the propagation of compressive
 slow mode MHD waves at smaller scales. It was further shown that the
 spectral behaviour is not influenced by the ambipolar diffusive
 process unlike viscous damping. Leake et al (2005) showed that the
 lower chromosphere contains neutral atoms, the existence of which
 greatly increases the efficiency of wave damping due to collisional
 friction momentum transfer.  They find that Alfv\'en waves with
 frequencies above 0.6Hz are completely damped and frequencies below
 0.01 Hz are unaffected. Khodachenko et al (2006) undertook a
 quantitative comparative study of the efficiency of the role of
 (ion-neutral) collisional friction, viscous and thermal conductivity
 mechanisms in damping MHD waves in different parts of the solar
 atmosphere.  It was pointed out by the authors that a correct
 description of MHD wave damping requires the consideration of all
 energy dissipation mechanisms through the inclusion of the
 appropriate terms in the generalized Ohm’s law, the momentum, energy
 and induction equations.  Molecular clouds are known to be supported,
 at least in part, by magnetic fields. The removal of magnetic fields
 thus represents an important component of the star formation
 process. In the most studied scenario, field removal occurs through
 the action of ambipolar diffusion, wherein magnetic fields are tied
 to the ionized component, which drifts relative to the more dominant
 neutral component of the gas (Mestel \& Spitzer 1956, Mouschovias
 1976, Nakano 1979, Shu 1983, Nakano 1984, Lizano \& Shu 1989, Basu \&
 Mouschovias 1994).  The role of magnetic fields and ion-neutral
 friction in regulating gravitationally driven fragmentation of
 molecular clouds was studied by Kudoh et al (2007).  Mestel \&
 Spitzer (1956) pointed out that even if clouds are magnetically
 supported, ambipolar diffusion (resulting from ion-neutral drag) will
 cause the support to be lost and stars to form.  Li \& Nakamura
 (2004) and Nakamura \& Li (2005) have studied turbulent fragmentation
 processes for a magnetized sheet including the effect of ion-neutral
 friction. They find that a mildly subcritical cloud can undergo
 locally rapid ambipolar diffusion and form multiple fragments because
 of an initial large scale highly supersonic compression wave.  Padoan
 et al (2000) calculated frictional heating by ion-neutral (or
 ambipolar) drift in turbulent magnetized molecular clouds and showed
 that the ambipolar heating rate per unit volume depends on field
 strength for constant rms Mach number of the flow, and on the
 Alfv\'enic Mach number. Furthermore, ion-neutral friction, has long
 been thought to be an important energy dissipation mechanism in
 molecular clouds, and therefore a significant heating mechanism for
 molecular cloud gas (Scalo 1977, Goldsmith \& Langer 1978, Zweibel \&
 Josafatsson 1983, Elmegreen 1985). These are only a few, amongst
 numerous other, studies that illustrate the importance of a neutral
 gas component in the dynamics of partially ionized magnetized
 astrophysical plasmas.

On even smaller scales, the edge region of a tokamak (a donut shaped
toroidal experimental device designed to achieve thermonuclear fusion
reaction in an extremely hot plasma) is partially ionized. In a
tokmak, the neutral particles result from effects such as gas puffing,
impurity injection, recombination, charge exchange, and possibly
neutral beam injection processes. The presence of neutrals can
potentially alter the dynamics of zonal flows and cross-field
diffusivity. For instance, several experiments have demonstrated that
the transition from L-mode (low confinement) to H-mode (high
confinement) can be significantly affected by neutral atoms in the
edge of a tokamak plasma \citep{singh2004}. The edge region
predominantly consists of neutral species resulting from recycling
from the wall, and from the limiter and/or divertor plate.  These
neutrals are present in significant numbers (e.g. $10^{10}- 10^{11}~
cm^{-3}$, which is about 5-8\% of the plasma density near the edge)
and affects the poloidal momentum balance and the L-H transition
threshold.  Thus they a play crucial role in regulating transport
processes in tokamak fusion plasmas.  The neutral gas can interact
with the magnetized plasma via charge exchange and other effects.
Furthermore, Doublet III-D tokamak (DIII-D) data shows that there is a
significant correlation between the power threshold for the L-H
transition and the poloidally averaged neutral density at the 95\%
flux surface \citep{groebner2001}. To understand the role of neutral
gas on H-mode physics, asymmetric gas puffing experiments have also
been performed in COMPASS-D and MAST devices
\citep{valovie2001,valovie2002}.

In all these environments, turbulence is ubiquitous. Turbulence
involves the nonlinear coupling and transfer of energy across a
multitude of spatial and temporal scales. The coupling of plasma to a
neutral gas via charge exchange, especially in a non-equilibrated
environment, introduces both a length scale distinct from the usual
collisional mfp and alternate channels and mode couplings for the
transfer of energy. Certainly, in the local ISM, the charge exchange
cross-section of e.g. neutral hydrogen atoms is larger than the
collisional cross-section (e.g. Zank 1999), implying that the former
governs much of the small-scale physics. Fluctuations with length
scales that exceed the charge exchange mean free path $\ell_{ce}$ will
obviously be effected by the neutral gas, such as the damping of
linear modes (Shaikh \& Zank 2006), but it is less clear how turbulent
plasma motion and nonlinear couplings will be mediated. For plasma
fluctuations on scales smaller than $\ell_{ce}$, the role of neutral
gas is even less clear.

Further, it is worth mentioning that the charge exchange process in
its simpler (and leading order) form can be treated like a friction or
viscous drag term in the fluid momentum equation, describing the
relative difference in the ion and neutral fluid velocities. The drag
imparted in this manner by a collision between ion and neutral also
causes ambipolar diffusion, a mechanism used to describe the Alfv\'en
wave damping by cosmic rays (Kulsrud \& Pearce 1969) and also
discussed by Oishi \& Mac Low (2006) in the context of molecular
clouds. The thermodynamical properties of the latter are significantly
different from that found in the heliosheath region. Interestingly,
unlike linear damping, ambipolar diffusion does not terminate an
isothermal MHD turbulence cascade (Oishi \& Mac Low 2006). Instead, it
damps some linear MHD waves. In the context of MHD turbulence, this
issue had also been addressed by Cho \& Lazarian (2003) who discussed
the viscosity-damped effects on inertial subrange associated with
magnetic field fluctuations in compressible (isothermal) MHD
turbulence. Ion-neutral friction nonetheless describes a leading order
interaction between the ion and neutral fluid species where thermal
speeds associated with both the fluids are of little significance or
ignorable and plasma phenomena are describable predominantly by
isothermal processes. This description is not appropriate to describe
the outer heliosheath plasma where proton as well as neutral hydrogen
temperatures can be as high as $20,000~^oK$. The thermal speeds
associated with the heliosheath protons and neutrals corresponding to
such a high temperature cannot therefore be ignored for the following
reasons. Firstly, since the thermal speed describes a significant
fraction of the energy and is proportional to the relative speeds in
the momentum and energy charge exchange sources, it should not be
neglected for plasmas that are not cold. It can potentially influence
the momentum and energy sources during the course of evolution. This
means that the interaction between the heliosheath protons and neutral
atoms is not mediated only by the corresponding bulk fluid speeds, but
it also depends critically on their thermal speeds.  Secondly, for
elastic and charge exchange collisions, the initial particle
distribution can be well described by Maxwellians even with different
streaming velocities and temperatures. Higher order effects are
introduced in the dynamics because the distributions are not really
Maxwellians due to prior collisions as pointed out by McNutt et
al. (1998).  The latter introduces Navier-Stokes-like corrections
(viscosity and thermal conductivity like terms) to the neutral
component of the equation. A consistent treatment of such processes,
describing the higher order corrections, in the context of the
heliosheath requires that the Maxwellians be multiplied by transfer
integrals, as done in Pauls et al (1995).  Thirdly, plasma and neutral
fluids in the heliosheath region are compressible enough to possess a
finite component of thermal energy associated with a non-adiabatic
exchange of energy amongst the local fluctuations.  Thus in the
context of supersonic heliosheath turbulent fluctuations, the
characteristic speed of plasma-neutral coupled fluid needs to be
treated on equal footing with the thermal speed. Consequently, thermal
corrections appear as higher order terms along with the leading order
ambipolar-like diffusive forces in the charge exchange expressions as
described in section 3.  This will be evident from the full energy
equation and the subsequent charge exchange energy transfer functions
employed in our model.

Multiple processes can couple the interstellar medium plasma and
neutral gas. Besides charge exchange scattering due to an induced
dipole moment in the neutral atom, depending on the temperature
(Banks, 1966), is possible. Interestingly enough, variations in
cross-sections can also be expected for different ions in the same
neutral gas at higher temperatures (Banks, 1966).  Further processes,
especially in the context of heliospheric processes, are
photoionization, electron impact ionization, recombination, and
H-H$^+$ Coulomb collisions (see e.g. Zank, 1999).  However,
cross-section associated with most of them is small compared to that
of charge exchange between hydrogen atom and ions in the heliosheath
region. We therefore do not incorporate them in our current model.

\begin{figure}
\plotone{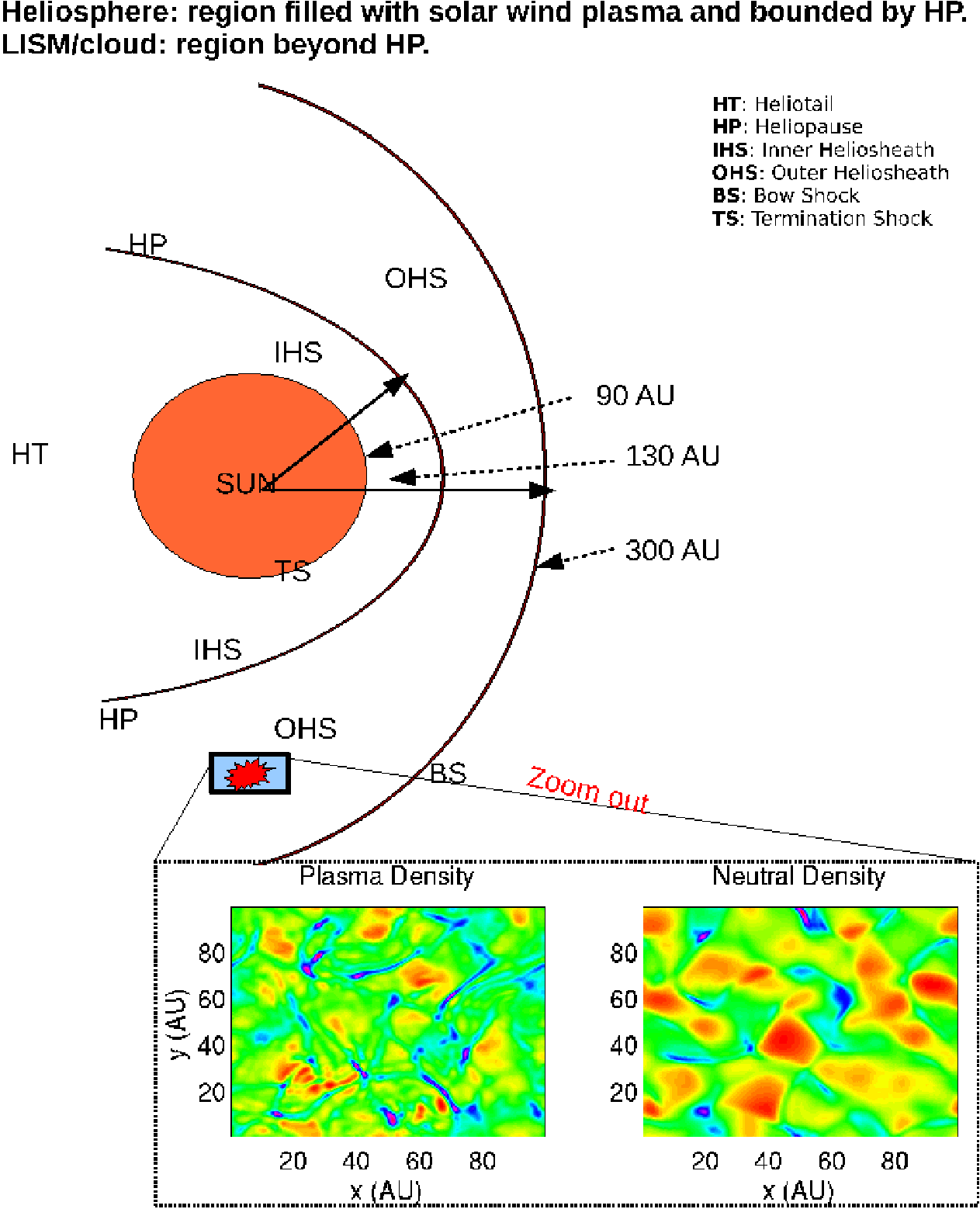} \figcaption{\label{fig0} Schematic illustrating
  the plasma neutral simulation regions both in the local interstellar
  medium and in the outer heliosheath (see \cite{zank1999} for a
  review of the solar wind interaction with the local ISM).}
\end{figure}

In view of these comments, we have developed a self-consistent
plasma-neutral ISM turbulence simulation model based on fluid modeling
of neutral and plasma fluids \citep{zank1999,pauls1995,zank1996b}. Our
model treats the plasma and neutral species as distinguishable fluids
and couples them through charge exchange.  The model essentially
employs a set of magnetohydrodynamic equations that describe the
turbulent plasma motions, whereas the neutrals are described by
hydrodynamic equations. A fluid approach for the local interstellar
medium is justifiable since the charge exchange and collisional mean
free path ensures approximate thermodynamical equipartition between
plasma and neutral species with the result that both plasma and
neutrals can be described by a Maxwellian distribution. Besides the
local ISM, the outer heliosheath can also be described reasonably well
by a fluid model of this kind, as was done in the model of
\citet{pauls1995}. For a non-equilibrated description, a more
elaborate multi-fluid model can be developed, as was done by
\citet{zank1996b,zank1996}, and this could be applied to
straightforwardly to both the outer heliosheath and very local
ISM. This would be more important to the outer heliosheath since a
multi-fluid model would then include hot inner heliosheath neutral
atoms deposited through secondary charge exchange in the outer
heliosheath \citep{zank1996b,zank1996,zank1999}.  The schematic in
\Fig{fig0} depicts the simulation regions in which we are particularly
interested i.e. the outer heliosheath and the local interstellar
medium. The importance of ISM turbulence is clearly evident from this
figure in the context of cosmic ray modulation, large-scale
heliospheric structure and particle acceleration.

In this paper, we focus on the multi-scale evolution of a partially
ionized plasma, comprising multiple fluids, where each fluid evolves
under the influence of the other through a complex interaction
process.  We investigate the energy cascade between LISM
turbulent-fluctuations in a partially ionized plasma where the plasma
and neutral gas are coupled through charge exchange. Here, we, present
a numerical simulation of ISM turbulence that self-consistently
evolves both the plasma and neutral fluids when coupled by charge
exchange. We restrict our attention to a neutral hydrogen (H) gas, so
that plasma and neutral densities are conserved by charge exchange
interactions.  Hence the corresponding continuity equations do not
include charge exchange terms.  One of the points that emerges from
our coupled multi fluid LISM simulations is that neutrals enhance
turbulent cascade rates and lead to much steeper spectra compared to
that predicted by the usual Kolmogorov theory \citep{k41} for pure MHD
or hydrodynamics \citep{shaikh2007}.  In Section 2, we discuss the
equations of a coupled plasma-neutral ISM turbulence model, their
validity, the underlying assumptions and the normalizations. Section 3
deals with a quantitative derivation of charge exchange source terms,
associated with the plasma and neutral fluids, based on the
time-dependent Boltzmann equation. Section 4 describes the results of
our nonlinear, coupled, self-consistent ISM fluid simulations. It is
shown, in section 5, that the coupling of plasma and neutral fluids
leads to an enhanced spectral transfer in the inertial range, which
results in a steeper ISM turbulent spectrum than predicted by the
Kolmogorov theory.  Finally, conclusions are presented in Section 6.

\section{Model Equations for a partially ionized  LISM}
The assumptions that are intrinsic to our model of turbulence in a
partially ionized ISM are the following.  (i) Fluctuations in the
plasma and neutral fluids are isotropic, homogeneous, thermally
equilibrated and turbulent, and (ii) Neither a mean magnetic field nor
velocity flows are present initially. Local mean flows may
subsequently be generated by self-consistently excited nonlinear
instabilities. (iii) The characteristic turbulent correlation
length-scales ($\lambda_c \sim 1/k_c$) are typically bigger than
charge-exchange mean free path lengths ($\lambda_{ce}\sim 1/k_{ce}$)
in the ISM flows, i.e $\lambda_c\gg \lambda_{ce}$ or $k_{ce}/k_c \gg
1$. The latter inequality is also consistent with
\citet{Florinski2003,Florinski2005}.  Nevertheless, they are large
enough to treat any localized shocks as smooth discontinuities. In
other words, the characteristic shock length-scales are small compared
to the ISM turbulent fluctuation length-scales, and finally (iv)
boundary conditions are periodic, essentially a box of interstellar
plasma, which is a natural and most appropriate choice for modeling
turbulence in the local ISM.

While most of the above assumptions are appropriate to realistic ISM
turbulent flows, so allowing us to use MHD and hydrodynamic
descriptions for the plasma and the neutral components respectively,
the latter has been criticized in the context of heliospheric flows
within the heliopause owing to the large charge-exchange mean free
path (about the size or even bigger than the extent of the
heliosphere).  The use of a fluid description for neutrals in the
inner heliosheath is therefore inappropriate.  In the LISM and outer
heliosheath, we are nonetheless not restricted by a large mfp because
the plasma and neutral fluid remain close to thermal equilibirium and
behave as Maxwellian fluids.  Our model simulates the local ISM and
outer heliosheath. The fluid model describing nonlinear turbulent
processes in the interstellar medium, in the presence of charge
exchange, can be cast into plasma density ($\rho_p$), velocity (${\bf
U}_p$), magnetic field (${\bf B}$), pressure ($P_p$) components
according to the conservative form
\be
\label{mhd}
 \frac{\partial {\bf F}_p}{\partial t} + \nabla \cdot {\bf Q}_p={\cal Q}_{p,n},
\ee
where,
\[{\bf F}_p=
\left[ 
\begin{array}{c}
\rho_p  \\
\rho_p {\bf U}_p  \\
{\bf B} \\
e_p
  \end{array}
\right], 
{\bf Q}_p=
\left[ 
\begin{array}{c}
\rho_p {\bf U}_p  \\
\rho_p {\bf U}_p {\bf U}_p+ \frac{P_p}{\gamma-1}+\frac{B^2}{8\pi}-{\bf B}{\bf B} \\
{\bf U}_p{\bf B} -{\bf B}{\bf U}_p\\
e_p{\bf U}_p
-{\bf B}({\bf U}_p \cdot {\bf B})
  \end{array}
\right],\\
{\cal Q}_{p,n}=
\left[ 
\begin{array}{c}
0  \\
{\bf Q}_M({\bf U}_p,{\bf V}_n, \rho_p, \rho_n, T_n, T_p)   \\
0 \\
Q_E({\bf U}_p,{\bf V}_n,\rho_p, \rho_n, T_n, T_p)
  \end{array}
\right]
\] 
and
\[ e_p=\frac{1}{2}\rho_p U_p^2 + \frac{P_p}{\gamma-1}+\frac{B^2}{8\pi}.\]
The above set of plasma equations is supplemented by $\nabla \cdot {\bf
B}=0$ and is coupled self-consistently to the ISM neutral density
($\rho_n$), velocity (${\bf V}_n$) and pressure ($P_n$) through a set
of hydrodynamic fluid equations,
\be
\label{hd}
 \frac{\partial {\bf F}_n}{\partial t} + \nabla \cdot {\bf Q}_n={\cal Q}_{n,p},
\ee
where,
\[{\bf F}_n=
\left[ 
\begin{array}{c}
\rho_n  \\
\rho_n {\bf V}_n  \\
e_n
  \end{array}
\right], 
{\bf Q}_n=
\left[ 
\begin{array}{c}
\rho_n {\bf V}_n  \\
\rho_n {\bf V}_n {\bf V}_n+ \frac{P_n}{\gamma-1} \\
e_n{\bf V}_n
  \end{array}
\right],\\
{\cal Q}_{n,p}=
\left[ 
\begin{array}{c}
0  \\
{\bf Q}_M({\bf V}_n,{\bf U}_p, \rho_p, \rho_n, T_n, T_p)   \\
Q_E({\bf V}_n,{\bf U}_p,\rho_p, \rho_n, T_n, T_p)
  \end{array}
\right],
\] 
\[e_n= \frac{1}{2}\rho_n V_n^2 + \frac{P_n}{\gamma-1}.\]

Equations (\ref{mhd}) to (\ref{hd}) form an entirely self-consistent
description of the coupled ISM plasma-neutral turbulent fluid.
Several points are worth noting.  The charge-exchange momentum sources
in the plasma and the neutral fluids, i.e. Eqs. (\ref{mhd}) and
(\ref{hd}), are described respectively by terms ${\bf Q}_M({\bf
  U}_p,{\bf V}_n,\rho_p, \rho_n, T_n, T_p)$ and ${\bf Q}_M({\bf
  V}_n,{\bf U}_p,\rho_p, \rho_n, T_n, T_p)$. A swapping of the plasma
and the neutral fluid velocities in this representation corresponds,
for instance, to momentum changes (i.e. gain or loss) in the plasma
fluid as a result of charge exchange with the ISM neutral atoms
(i.e. ${\bf Q}_M({\bf U}_p,{\bf V}_n,\rho_p, \rho_n, T_n, T_p)$ in
Eq. (\ref{mhd})). Similarly, momentum change in the neutral fluid by
virtue of charge exchange with the plasma ions is indicated by ${\bf
  Q}_M({\bf V}_n,{\bf U}_p,\rho_p, \rho_n, T_n, T_p)$ in
Eq. (\ref{hd}). In the absence of charge exchange interactions, the
plasma and the neutral fluid are de-coupled trivially and behave as
ideal fluids.  While the charge-exchange interactions modify the
momentum and the energy of plasma and the neutral fluids, they
conserve density in both the fluids (since we neglect photoionization
and recombination). Nonetheless, the volume integrated energy and the
density of the entire coupled system will remain conserved in a
statistical manner. The conservation processes can however be altered
dramatically in the presence of any external forces. These can include
large-scale random driving of turbulence due to any external forces or
instabilities, supernova explosions, stellar winds, etc. Finally, the
magnetic field evolution is governed by the usual induction equation,
i.e. Eq. (\ref{mhd}), that obeys the frozen-in-field theorem unless
some nonlinear dissipative mechanism introduces small-scale damping.

The underlying ISM turbulence model can be non-dimensionalized
straightforwardly using a typical ISM scale-length ($\ell_0$), density
($\rho_0$) and velocity ($v_0$). The normalized plasma density,
velocity, energy and the magnetic field are respectively;
$\bar{\rho}_p = \rho_p/\rho_0, \bar{\bf U}_p={\bf U}_p/v_0,
\bar{P}_p=P_p/\rho_0v_0^2, \bar{\bf B}={\bf B}/v_0\sqrt{\rho_0}$. The
corresponding neutral fluid quantities are $\bar{\rho}_n =
\rho_n/\rho_0, \bar{\bf U}_n={\bf U}_n/v_0,
\bar{P}_n=P_n/\rho_0v_0^2$. The momentum and the energy
charge-exchange terms, in the normalized form, are respectively
$\bar{\bf Q}_m={\bf Q}_m \ell_0/\rho_0v_0^2, \bar{Q}_e=Q_e
\ell_0/\rho_0v_0^3$. The non-dimensional temporal and spatial
length-scales are $\bar{t}=tv_0/\ell_0, \bar{\bf x}={\bf
  x}/\ell_0$. Note that we have removed bars from the set of
normalized coupled ISM model equations (\ref{mhd}) \& (\ref{hd}).  The
charge-exchange cross-section parameter ($\sigma$), which does not
appear directly in the above set of equations (see the subsequent
section for more detail), is normalized as $\bar{\sigma}=n_0 \ell_0
\sigma$, where the factor $n_0\ell_0$ has dimension of (area)$^{-1}$.
By defining $n_0, \ell_0$ through
$\sigma_{ce}=1/n_0\ell_0=k_{ce}^{-2}$, we see that there exists a
critical charge exchange wavenumber ($k_{ce}$) associated with the
coupled ISM plasma-neutral turbulent system.  For a characteristic
density, this corresponds physically to an area defined by the charge
exchange mode being equal to (mfp)$^2$.  Thus the larger the area, the
higher is the probability of charge exchange between plasma ions and
neutral atoms, as illustrated in Fig. (2).  To be precise, `$k_{ce}$'
is a (lengthscale)$^{-1}$ that typically helps us determine whether or
not a particular turbulent fluctuation length scale undergoes charge
exchange in our model. It is strictly in that sense, we refer to it as
a critical wavenumber. Clearly, there exist three different regimes
depending on whether (i) $k<k_{ce}$, (ii) $k \simeq k_{ce}$ or (iii)
$k>k_{ce}$. In the heliosheath, the probability that charge exchange
can directly modify those modes satisfying $k<k_{ce}$ is high compared
to modes satisfying $k>k_{ce}$.  Since the charge exchange
length-scales are much smaller than the ISM turbulent correlation
scales, this further allows many charge exchange interactions amongst
the nonlinear turbulent ISM modes before they cascade energy in one
unit convective time. This is illustrated schematically in Fig. (2).

It is interesting to compare our model with other work. Of particular
relevance is the classic work by Kulsrud and Pearce (1969) who
investigated the interaction of galactic comic rays and Alfi\'en waves
in the interstellar medium and also considered ion-neutral collision
as a damping mechanism for Alfv\'en waves.  Ion-neutral collisions are
also reported by McIvor (1977) to be a dominant mechanism for damping
interstellar medium turbulence.  The damping (due to ion-neutral
collisions) rates for Alfv\'en waves estimated by Kulsrud and Pearce
(1969) are based on $\nu_0 \simeq n_0 v_{th} \sigma$, where $\nu_0,
n_0, v_{th}$ and $\sigma$ are respectively the ion-neutral collision
frequency, neutral density, thermal speed and collision cross-section.
The damping wavenumber associated with this frequency can be estimated
to be $k\simeq n_0 v_{th} \sigma/V_A$. The latter determines
essentially the dissipation wavenumber associated with the damping of
small scale Alfv\'en waves in ISM turbulence (Kulsrud and Pearce
1969). The turbulent cascade is further expected to be terminated
beyond this wavenumber. It is clear from this expression that the
higher the thermal speed, the larger the dissipation
wavenumber. Correspondingly in real space, dissipation will be
concentrated on smaller length-scales for higher thermal
speeds. Furthermore, the thermal speed is also proportional to the
temperature.  Thus, given the heliosheath parameters, $v_{Thu}$
associated with {\it heliosheath} neutral atoms will be at least two
orders of magnitude larger compared with that results from Eq. (47) in
Kulsrud and Pearce (1969).  This will correspond to a larger
dissipative wavenumber in the heliosheath region and essentially means
that charge exchange interactions are restricted to only those eddies
whose length-scales are comparable to the dissipative length-scales of
turbulence. It is not clear if this scenario holds in inner/outer
heliosheath turbulence where charge exchange interactions, in
principle, can modify any length scale in the inertial range.  Part of
the reason resides with the fact that charge exchange modifications
are proportional to the relative speed between ion and neutral fluids
and corresponding densities which can obviously alter the nonlinear
cascades by influencing the combined momenta and energies of both the
fluids in a self-consistent and subtle manner. Additionally, Kulsrud
and Pearce (1969), as does McIvor (1977), assumed a fixed neutral
density while computing the damping rates associated with Alfv\'en
waves in ISM turbulence. While a fixed neutral density can modify the
linear growth rates of the underlying waves, it does not itself evolve
and hence no back reaction on Alfv\'en waves and vice versa can be
expected.  By contrast, our model self-consistently evolves neutral
and plasma and describes the mutual feedback of one species on the
other.  This not only alters the linear interaction process, but it
also influences the nonlinear coupling in a subtle manner not readily
describable by means of any linear analytic theory.  A self-consistent
coupling is absolutely essential to understand nonlinear turbulent
cascades in the inner/outer heliosheath. It is to be noted further
that we do not attempt to investigate damping of heliosheath
turbulence by Alfv\'en waves that experience ion-neutral collisions,
unlike McIvor (1977).  While Alfv\'en waves are intrinsically present
in our model, turbulent cascades are least affected by them along the
mean magnetic field (Shebalin \& Montgomery 1983). Alfv\'en waves can
be crucial in providing a local anisotropy in the spectral transfer
due to a mean or local field though.  However, our prime focus here is
to develop an understanding of heliosheath turbulence where plasma and
neutral fluids evolve through a mutual interaction mediated by charge
exchange. As evident from the complexity associated with these
interactions, nonlinear turbulent cascades are likely to be affected
at any length scale in the heliosheath inertial range, unlike a linear
dissipative processes.

In the subsequent section, we derive an exact quantitative form of the
sources due to charge exchange in ISM turbulence.

\section{Charge Exchange Sources}
The charge exchange terms can be obtained from the Boltzmann transport
equation that describes the evolution of a neutral distribution
function $f_n=f({\bf x}, v_x, v_y, v_z, t)$ in a six-dimensional phase
space defined respectively by position and velocity vectors $({\bf x},
v_x, v_y, v_z)$ at each time $t$. Here we follow \citet{pauls1995} in
computing the charge exchange terms from various moments of the
Boltzmann equation. The Boltzmann equation for the neutral
distribution contains a source term proportional to the proton
distribution function $f_p$ and a loss term proportional to the
neutral distribution function $f_n$, \eqa
\label{bol}
\frac{\partial f_n}{\partial t} + {\bf v}_n \cdot \nabla f_n +
\frac{\bf F}{m} \cdot \nabla_{{\bf v}_n} f_n = 
f_{p}({\bf x}, v_x, v_y, v_z, t) \int f_n({\bf x}, v_x, v_y, v_z, t) |{ v}_n- {u}_p|
\sig(v_{rel}) d^3{v}_n \nonumber \\ 
-f_n({\bf x}, v_x, v_y, v_z, t) \int
f_p({\bf x}, v_x, v_y, v_z, t) |{ u}_p- { v}_n| \sig(v_{rel})
d^3{ u}_p.  \eeq Here $f_p, ~{u}_p$ represent respectively the
proton distribution function and velocity.  $\sig$ is the charge
exchange cross-section (between neutrals and plasma protons), $m$ is
the mass of particle, and ${\bf F}$ represents forces acting on the
fluid.  The charge exchange parameter has a logarithmically weak
dependence on the relative speed ($v_{rel}=|{u}_p- {v}_n|$) of
the neutrals and the protons through $\sig = [(2.1-0.092 \ln
  (v_{rel})) 10^{-7} cm]^2$ \citep{fite}.  This cross-section is valid
as long as energy does not exceed $1eV$, which usually is the case in
the inner/outer heliosphere. Beyond $1eV$ energy, this cross-section
yields a higher neutral density. This issue is not applicable to our
model and hence we will not consider it here.  The density, momentum,
and energy of the thermally equilibrated Maxwellian proton and neutral
fluids can be computed from \eq{bol} by using the zeroth, first and
second moments $\int f_{\xi} d^3\xi, \int m{\bf \xi} f_{\xi} d^3\xi$
and $\int m\xi^2/2 f_{\xi} d^3\xi$ respectively, where $\xi={u}_p$
or ${v}_n$. Since charge exchange conserves the density of the
proton and neutral fluids, there are no sources in the corresponding
continuity equations. We, therefore, need not compute the zeroth
moment of the distribution function.  Computing directly the first
moment from \eq{bol}, we obtain the neutral fluid momentum equation as
given by \eq{hd}.  The entire rhs of \eq{bol} can now be replaced by a
momentum transfer function ${\bf Q}_M({v}_n,{u}_p) $ which
reads \be
\label{qm}
 {\bf Q}_M({v}_n,{u}_p) = \bar{\mu}({u}_p,{v}_n)-\bar{\mu}({v}_n,{u}_p),
\ee
where ${\bf Q}_M$ and $ \bar{\mu}$, the transfer integral, are 
vector quantities. The transfer integrals describe  the
charge exchange transfer of momentum from proton to neutral fluid and
vice versa.  The first term on the rhs of
Eq. (4) can be expressed by
\[\bar{\mu}({u}_p,{v}_n) = f_p({\bf x}, v_x,v_y,v_z,t)\beta({u}_p,{v}_n)\]
where,
\[\beta({u}_p,{v}_n)=\int f_n({\bf x}, v_x,v_y,v_z,t) |{v}_n- {u}_p|
\sig(v_{rel}) d^3{v}_n.\]
Considering a Maxwellian distribution for the neutral atoms, we simplify 
$\beta({u}_p,{v}_n)$ as
follows,
\[\beta({u}_p,{v}_n)= \sig n_n V_{T_n} 
\sqrt{\frac{4}{\pi}+\frac{({v}_n- {u}_p)^2}{V_{T_n}^2}}.\] Note that
the above expression emerges directly from a straightforward
integration of sources in the rhs of the Boltzmann Eq. (3). To obtain
the expression for the momentum transferred from proton to neutral (or
vice versa), we need to take a second moment of the
$\bar{\mu}({u}_p,{v}_n)$ expression. This is shown in the following,
\[\bar{\mu}({u}_p,{v}_n) = m {\bf v}_n I_0({u}_p,{v}_n) + m ({\bf u}_p-{\bf v}_n) I_1({u}_p,{v}_n).\]
where $I_0$ and $I_1$ are transfer integrals that can be written as follows,
\[I_0({u}_p,{v}_n)= \int  f_p({\bf x}, v_x,v_y,v_z,t)\beta({u}_p,{v}_n) ~d^3{u}_p;\]
\[I_1({u}_p,{v}_n)= \int {v}_n f_p({\bf x}, v_x,v_y,v_z,t)\beta({u}_p,{v}_n) ~d^3{u}_p.\]
Assuming a Maxwellian distribution for plasma protons and using
$\beta({u}_p,{v}_n)$ from the above expression, we can
straightforwardly evaluate the transfer integrals $I_0$ and $I_1$ (see
Pauls et al (1995) for details).  We further write the
complete form of the first term on the rhs of Eq. (4) as follows,
\[\bar{\mu}({u}_p,{v}_n) = m\sig n_p n_n \left[U_{{u}_p,{v}_n}^\ast{\bf v}_n - ({\bf u}_p-{\bf v}_n)
\frac{  V_{T_n}^2}{\delta V_{{u}_p,{v}_n}} \right].\] In a similar
manner, we can evaluate the second term on the rhs of Eq. (4), which
yields the following form,
\[\bar{\mu}({v}_n,{u}_p) = m\sig n_{p}n_n \left[U_{{v}_n,{u}_p}^\ast {\bf u}_p - ({\bf v}_n-{\bf u}_p) \frac{V_{T_p}^2}{\delta V_{{v}_n,{u}_p}} \right],\]
where $\delta V_{{u}_p,{v}_n} =
  \left[4\left(\frac{4}{\pi}V_{T_p}^2+\Delta U^2 \right)
  +\frac{9\pi}{4}V_{T_n}^2 \right]^{1/2}, \delta V_{{v}_n,{u}_p} = \left[4\left(\frac{4}{\pi}V_{T_n}^2+\Delta U^2 \right)
  +\frac{9\pi}{4}V_{T_{n}}^2 \right]^{1/2}$ and $U^\ast=U_{{u}_p,{v}_n}^\ast = U_{{v}_n,{u}_p}^\ast =
  \left[\frac{4}{\pi}V_{T_{p}}^2+\frac{4}{\pi}V_{T_{n}}^2 +\Delta
  U^2\right]^{1/2}, \Delta {u} = {u}_p-{v}_n$ \citep{pauls1995}.  On
  substituting these expressions in the momentum transfer function, we
  obtain 
\be
\label{qm2}
 {\bf Q}_M({v}_n,{u}_p) = m\sig n_{p}n_n ({\bf v}_n -{\bf
   u}_p) \left[ U^\ast + \frac{V_{T_n}^2}{\delta V_{{u}_p,{v}_n}} -\frac{V_{T_p}^2}{\delta V_{{v}_n,{u}_p}}
   \right].  
\ee 
\Eq{qm2} together with \eq{hd} yields the momentum equation for
the neutral gas. Swapping the plasma and neutral fluid velocities
yields the corresponding source term for the proton fluid momentum
equation.  The gain or the loss in neutral or proton fluid momentum
depends upon the charge exchange sources, which depend upon the
relative speeds between neutrals and the protons. The thermal speeds
of proton and neutral gas in \eq{qm2} are given respectively by $
V_{T_p}^2 = 2K_BT_{p}/m$ (the factor 2 arises because of thermal
equilibration in that it is assumed that the temperature of the plasma
electrons and protons are nearly identical so that $T_p = T_e + T_{\rm
  proton} \simeq 2T_p$) and $V_{T_n}^2 = K_BT_{n}/m$. The
corresponding temperatures are related to the pressures by $
P_{p}=2n_{p}K_BT_{p}$ and $P_{n}=n_{n}K_BT_{n}$, where $n_n, T_n,
n_{p}, T_{p}$ are respectively the neutral and plasma density and the
temperature, and $K_B$ is the Boltzmann constant.

The  moment, $\int m\xi^2/2 f_{\xi} d^3\xi$, of the Boltzmann
\eq{bol} yields an energy equation for the neutral fluid whose
rhs contains the charge exchange energy transfer function 
\[Q_E({v}_n,{u}_p) = \eta({u}_p,{v}_n)-\eta({v}_n,{u}_p),\]
where $\eta({u}_p,{v}_n), ~\eta({v}_n,{u}_p)$ are the
energy transfer (from neutral to proton and vice versa) rates. These
functions can be computed as follows: 
\[\eta({u}_p,{v}_n) = \frac{1}{2}
mV_n^2 \sig n_{p}n_n U^\ast + \frac{3}{4}m V_{T_n}^2 \sig n_{p}n_n \Delta
V_{{u}_p,{v}_n}- m \sig n_pn_n {\bf v}_n\cdot ({\bf u}_p-{\bf
v}_n) \frac{V_{T_n}^2}{\delta V_{{u}_p,{v}_n}}\]
 and 
\[\eta({v}_n,{u}_p) = \frac{1}{2}mV_{n}^2 \sigma n_{p}n_n U^\ast +  \frac{3}{4}m
V_{T_{p}}^2 \sig n_{p}n_n \Delta V_{{v}_n,{u}_p}- m \sig
n_{p}n_n {\bf u}_p \cdot ({\bf v}_n-{\bf u}_p) \frac{V_{T_{p}^2}}{\delta
V_{{v}_n,{u}_p}}.\]  
The total energy transfer from neutral to proton fluid, due to charge
exchange, can then be written as,
\eqa
\label{qe}
Q_E({v}_n,{u}_p) = \frac{1}{2}m \sig  n_{p}n_n U^\ast (V_n^2-U_p^2)+
\frac{3}{4}m  \sig  n_{p}n_n (V_{T_{n}}^2\Delta V_{{u}_p,{v}_n}-V_{T_{p}}^2
\Delta V_{{v}_n,{u}_p}) 
 \nonumber \\
-m \sig  n_{p}n_n
\left[ {\bf v}_n \cdot ({\bf u}_p-{\bf v}_n)\frac{V_{T_n}^2}{\delta V_{{u}_p,{v}_n}}-
{\bf u}_p \cdot ({\bf v}_n-{\bf u}_p)\frac{V_{T_{p}^2}}{\delta V_{{v}_n,{u}_p}}\right],
\eeq
with $\Delta V_{{u}_p,{v}_n} = 
[\frac{4}{\pi}V_{T_p}^2+\frac{64}{9\pi}V_{T_n}^2 +\Delta U^2]^{1/2} ~{\rm and} ~
\Delta V_{{v}_n,{u}_p} = [\frac{4}{\pi}V_{T_n}^2+
\frac{64}{9\pi}V_{T_{p}}^2 +\Delta U^2]^{1/2}$.
A similar expression for the energy transfer charge exchange source
term of plasma energy in \eq{mhd} can be obtained by exchanging the
plasma and neutral fluid velocities. In the normalized momentum and
energy charge exchange source terms, the factor $m\sig$ in
\eqs{qm2}{qe} is simply replaced by $\bar{\sigma}$.

\section{ Nonlinear Simulation of Energy Cascades}
It is clear from previous sections that the interaction of the
neutrals with a plasma introduces characteristic length and time scales
that can separate characteristic time and length scales in ISM
turbulence. For instance, the plasma-neutral fluid coupling in the ISM
fluctuations introduces charge exchange modes, $k_{ce}$, that are
distinctively different from the characteristic turbulent mode
$k$. Typically, characteristic turbulent length-scales ($\lambda_c\sim
1/k$) are much bigger than the charge exchange length-scales
$\lambda_{ce}\sim1/k_{ce}$, since for example, $\lambda_c \sim
(100-1000)\lambda_{ce}$ in the ISM cloud
\citep{Florinski2003,Florinski2005}. This essentially corresponds to
$k_{ce}/k\gg1$ \citep{zank1999,Florinski2003,Florinski2005}. It should
be noted that the charge exchange sources in the neutral and the
plasma fluids do not merely damp the smallest scale fluctuations, such
as is the case with linear collisional dissipation in turbulent fluid
flows, but they can alter the nonlinear cascade rate dramatically
\citep{shaikh2006,shaikh2007}. Because of its complex nonlinear
character (see \eqs{qm2}{qe}), charge exchange can be {\it effective on
  a variety of turbulent length-scales} unlike linear dissipation which
is operative on small length-scales only.  The effect and influence of
the charge exchange source terms on the turbulent inertial range of
coupled neutral and plasma fluids is therefore a subtle issue and is
not restricted only to the damping of small scales in partially
ionized, magnetized ISM turbulence. With this in mind, we carry out
two different sets of simulations of \eqs{mhd}{hd}, to discern the
effects of charge exchange on the ISM plasma-neutral system. In the
first set of simulations, \eqs{mhd}{hd} are decoupled (i.e. no charge
exchange coupling is assumed) and each evolves as an ideal fluid,
whereas in the second set the coupling is included self-consistently.
The former provides a reference simulation against which to
investigate the coupled system.

A two-dimensional (2D) nonlinear fluid code was developed to
numerically integrate \eqsto{mhd}{hd}. The 2D simulations are not only
computationally less expensive (compared to a fully 3D calculation),
but they offer significantly higher resolution (to compute inertial
range turbulence spectra) even on moderately-sized clusters like our
local Beowulf.  The spatial discretization in our code uses a discrete
Fourier representation of turbulent fluctuations based on a
pseudospectral method, while we use a Runge Kutta 4 method for the
temporal integration. All the fluctuations are initialized
isotropically (no mean fields are assumed) with random phases and
amplitudes in Fourier space.  This algorithm ensures conservation of
total energy and mean fluid density per unit time in the absence of
charge exchange and external random forcing. Additionally, $\nabla
\cdot {\bf B}=0$ is satisfied at each time step.  Our code is
massively parallelized using Message Passing Interface (MPI) libraries
to facilitate higher resolution. The initial isotropic turbulent
spectrum of fluctuations is chosen to be close to $k^{-2}$ with random
phases in both $x$ and $y$ directions.  The choice of such (or even a
flatter than -2) spectrum does not influence the dynamical evolution
as the final state in our simulations progresses towards fully
developed turbulence.  While the ISM turbulence code is evolved with
time steps resolved self-consistently by the coupled fluid motions,
the nonlinear interaction time scales associated with the plasma
$1/{\bf k} \cdot {\bf U}_p({\bf k})$ and the neutral $1/{\bf k} \cdot
{\bf V}_n({\bf k})$ fluids can obviously be disparate. Accordingly,
turbulent transport of energy in the plasma and the neutral ISM fluids
takes place on distinctively separate time scales.

Because of the different nonlinear time scales associated with the
plasma and neutral fluids, mode structures can be different in the two
fluids when they are evolved together and in isolation
(i.e. decoupled). The former is shown in \fig{fig3}{fig4}. The figures
respectively show the evolution of various physical quantities in the
plasma-neutral coupled system for a typical set of ISM parameters as
described in the captions (for typical ISM parameters, see also
\cite{zank1999,zank1996b,jacob2006,Florinski2003,Florinski2005}).  In
the absence of charge exchange (see \Fig{fig3}, right column), the
plasma fluid evolves as an ideal MHD fluid and is known to typically
develop sheet-like structures in the magnetic field
\citep{biskamp}. By contrast, the evolution of the magnetic field is
modified substantially when the two fluids are coupled through charge
exchange. For instance, the sheet-like structures present in the
turbulent magnetic field of the uncoupled plasma system instead are
smeared in the coupled plasma-neutral system on the typical
sheet-formation time-scales of ideal MHD.  The neutral fluid, under
the action of charge exchange terms, tends to modify the cascade rates
by isotropizing the ISM turbulence on a slower time scale. This point
is further elucidated in the subsequent section based on
Kolmogorov-like phenomenological arguments.  However if the coupled
system is evolved further, sheets do form eventually in the magnetic
field (after a relatively long time compared to the ideal MHD
time-scales) and a turbulent equipartition is set up in the plasma and
the neutral fluid modes.  In any event, the small-scale sheet-like
structures in magnetic field compress (or pinch) the plasma
density. Hence the density fluctuations develop identical (thinner
than) sheet-like structures that co-exist with small-scale turbulent
fluctuations in their spectrum as shown in \Fig{fig3}. The neutral
fluid, on the other hand, evolves isotropically as stated above by
forming relatively large-scale structures (see \Fig{fig4}).

\section{Energy Spectra}
Spectral transfer in ISM turbulence progresses under the action of
nonlinear interactions as well as charge exchange sources. Energy
cascades amongst turbulent eddies of various scale sizes and between
the plasma and the neutral fluids. In a freely decaying case, plasma
and neutral fluids evolve under the influence of charge exchange
forces which dramatically affect the energy cascades in the inertial
range. This is evident from \fig{fig5}{fig6} where Kolmogorov-like
fully developed turbulent spectra in the inertial range are shown
respectively for the coupled neutral and plasma fluids.  The neutral
fluid energy spectrum in the inertial regime for the coupled system
exhibits a $k^{-\alpha}$ spectrum, where the spectral index $\alpha
\approx 3.2$. On the other hand, the spectral index for plasma
magnetic and kinetic energy spectra for the coupled system is $\alpha
\approx 2.4$.  The inertial range spectral indices for the coupled
plasma-neutral system in our simulations show a {\it significant}
deviation from their corresponding uncoupled analogues which are
respectively $-3$ and $-1.7$ for the neutral and plasma fluids, also
plotted in \fig{fig5}{fig6}. The inertial range turbulent spectra are
much steeper for the coupled ISM plasma-neutral system than the
corresponding uncoupled or isolated neutral and plasma fluids in the
decaying turbulence regime. Consequently, the steep turbulent spectra
in the coupled ISM plasma-neutral system gives rise to an enhanced
spectral transfer amongst the inertial range modes, and the energy
cascade rates associated with the plasma and neutral (coupled) fluids
can be understood by invoking a Kolmogorov-like phenomenology as
described in the following.

The typical nonlinear interaction time-scale in ordinary
(i.e. uncoupled) plasma and/or neutral turbulence is given by
\be
\tau_{nl} \sim \frac{\ell_0}{v_\ell} \sim (kv_k)^{-1},
\ee 
where $v_k$ or $v_\ell$ is the velocity of turbulent eddies. In the
presence of charge exchange interactions, the ordinary nonlinear
interaction time-scale of fluid turbulence is modified by a factor
$k_{ce}/k$ such that the new nonlinear interaction time-scale of ISM
turbulence is now 
\be
\tau_{NL} \sim \frac{k_{ce}}{k} \frac{1}{kv_k}.
\ee 
The main justification of using the above expression originates from
the fact that turbulent cascade is determined typically by the
nonlinear interaction time ($\tau_{nl} \sim 1/kv_k$, where $\tau_{nl}$
is the eddy interaction time, $k$ is characteristic turbulent
wavenumber, and $v_k$ is characteristic speed of eddy in wavenumber
space) associated with the convection of turbulent eddies in the fluid
momentum equation. Note that this time scale follows from Kolmogorov's
phenomenology (Kolmogorov 1941) and corresponds essentially to the
eddy interaction time in an ideal, non interacting, and non magnetized
hydrodynamic-like turbulent fluid. Such time scales are essential to
estimate spectral energy transfer (or cascades) rates ($\varepsilon
\simeq E(k)/\tau_{nl}$, where $E(k)$ is turbulent energy per unit
mode) in a hydrodynamic turbulent fluid. For a non-ideal, interacting
and magnetized fluid, incorporating complex interactions amongst the
turbulent eddies with a background or local mean magnetic field, the
nonlinear eddy interaction time needs to be modified. This was first
pointed out by Kraichnan (1965) in the context of magnetohydrodynamic
(MHD) fluids where the interaction of turbulent eddies with Alfv\'en
waves, excited as a result of a mean magnetic field, was
addressed. Owing to the presence of waves in a MHD fluid, turbulent
correlations between velocity and magnetic field and the corresponding
energy transfer time are determined primarily by $\tau \sim (b_0
k)^{-1}$, where $b_0$ is a typical amplitude of a local magnetic field
and is dimensionally identical to that of the velocity field by virtue
of Els$\ddot{\rm a}$sser's symmetry (Kraichnan 1965). Note carefully
the modification in the energy transfer (or cascade) time scale
because of the wave-turbulent eddy interaction process.  Since the
convective time is the only nonlinear time scale involved in the
spectral transfer of turbulent energy, it it this time that accounts
entirely for the nonlinear cascade in the coupled plasma-neutral
heliosheath turbulence.  Likewise, we introduce an analogous
modification in the coupled plasma-neutral ISM turbulence where the
energy cascade time scale due to charge exchange interactions amongst
turbulent eddies, is determined explicitly by the factor $k_{ce}/k$
(where the symbols have their usual meanings).  It is this factor that
determines how plasma and neutral fluids are coupled in heliosheath
turbulence. Accordingly, the nonlinear spectral transfer time or the
energy cascade rate is modified and is proportional to the factor
$k_{ce}/k$ such that the new nonlinear eddy interaction time in the
coupled plasma-neutral heliosheath turbulence is $\tau_{NL} \sim
(k_{ce}/k) (1/kv_k)$. The interaction time $\tau_{NL}$ obtained in
this manner for our coupled plasma-neutral problem not only adequately
reproduces all three different regimes of interactions, but it also
consistently describes the underlying physics in the context of
heliosheath turbulence.

On using the fact that $k_{ce}$ is typically larger than $k$ (or
$k_c$, the characteristic turbulent mode, as defined elsewhere in the
paper), i.e. $k_{ce}/k > 1$ in the ISM
\citep{Florinski2003,Florinski2005,zank1999}, the new nonlinear time
is $k_{ce}/k$ times bigger than the old nonlinear time i.e.
$\tau_{NL} \sim (k_{ce}/k) \tau_{nl}$. This enhanced nonlinear
interaction time in ISM turbulence is likely to prolong turbulent
energy cascade rates.  It is because of this enhanced or prolonged
interaction time that a relatively large spectral transfer of
turbulent modes tends to steepen the inertial range turbulent spectra
in both plasma and neutral fluids. By extending the above
phenomenological analysis, one can deduce exact (analytic) spectral
indices of the inertial range decaying turbulent spectra, as
follows. The new nonlinear interaction time-scale of ISM turbulence
can be rearranged as
\be
\tau_{NL} \sim
\frac{k_{ce}v_k}{kv_k}\frac{1}{kv_k}\sim \frac{\tau_{nl}^2}{\tau_{ce}},\ee 
where $\tau_{ce}\sim (k_{ce} v_k)^{-1}$ represents the charge exchange
time scale. The energy dissipation rate associated with the coupled
ISM plasma-neutral system can be determined from $\varepsilon \sim
E_k/\tau_{NL}$, which leads to 
\be
\varepsilon \sim \frac{v_k^2}{k_{ce}/k^2 v_k}\sim \frac{k^2 v_k^3}{k_{ce}}.\ee
According to the Kolmogorov theory, the spectral cascades are local in
$k$-space and the inertial range energy spectrum depends upon the
energy dissipation rates and the characteristic turbulent modes, such
that $E_k \sim \varepsilon^{\gamma} k^{\beta}$.  Upon substitution of
the above quantities and equating the power of identical bases, one
obtains
\be
E_k \sim \varepsilon^{2/3} k^{-7/3}\ee
for the plasma spectrum (the forward cascade inertial range). The spectral
index associated with this spectrum, i.e. $7/3 \approx 2.33$, is
consistent with the plasma spectrum observed in the (coupled ISM
plasma-neutral) simulations (see \fig{fig5}{fig6}). Similar arguments in the
context of neutral fluids, when coupled with the plasma fluid in ISM,
lead to the energy dissipation rates
 \be
\varepsilon \sim \frac{k^2 v_k^2}{k_{ce}/k^2 v_k}.\ee
 This further yields
 the forward cascade (neutral) energy spectrum 
\be
E_k \sim \varepsilon^{2/3} k^{-11/3}\ee
 which is close to the simulation result  shown in \Fig{fig6}.

A key issue finally is to understand what role charge exchange modes
play in coupled plasma-neutral ISM turbulence as they result
essentially from the nonlinear charge exchange interactions. To
address this issue, we plot charge exchange sources associated with
the momentum and energy equations, \eqs{mhd}{hd}, in \Fig{fig7}. It
appears from \Fig{fig7} that spectral energy is transferred
predominantly at the larger scales by means of charge exchange mode
coupling processes. The latter couples the large-scales, or smaller
than $k_{ce}$ modes, efficiently to low-$k$ ISM turbulent modes.  It
is primarily because of this $k_{ce}-k$ mode-coupling in the smaller
$k$ part of the spectrum of the coupled plasma-neutral ISM turbulence,
that energy is pumped efficiently at the lower $k$ inertial range
turbulent modes. The efficient coupling of the Fourier modes at low
$k$'s further enhances the nonlinear eddy time-scales associated with
the coupled plasma-neutral turbulence system which is consistent with
the scaling $\tau_{NL} \sim (k_{ce}/k) \tau_{nl}$, where $k_{ce}/k
>1$, as described above.  This consequently leads to the steepening of
the inertial range spectra observed in \fig{fig5}{fig6}. By contrast,
higher $k$ modes, far from the energy cascade inertial range, are
notably inefficient in transferring energy and momentum via charge
exchange mode coupling interactions and are damped by small-scale
dissipative processes in the coupled plasma-neutral ISM turbulence.

\section{Conclusions}
In conclusion, we have developed a self-consistent fluid model to
describe nonlinear turbulent processes in a partially ionized and
magnetized ISM gas. The charge exchange interactions couple the ISM
plasma and the neutral fluids by exciting a characteristic charge
exchange coupling mode $k_{ce}$, which is different from the
characteristic turbulent mode $k$ of the coupled system. One of the
most important points to emerge from our studies is that charge
exchange modes modify the ISM turbulence cascades dramatically by
enhancing nonlinear interaction time-scales on large scales.  Thus on
scales $\ell\ge\ell_{ce}$, the coupled plasma system evolves
differently than the uncoupled system where large-scale turbulent
fluctuations are strongly correlated with charge-exchange modes and
they efficiently behave as driven (by charge exchange) energy
containing modes of ISM turbulence.  By contrast, small scale
turbulent fluctuations are unaffected by charge exchange modes which
evolve like the uncoupled system as the latter becomes less important
near the larger $k$ part of the ISM turbulent spectrum (see also
\Fig{fig7}).  The neutral fluid, under the action of charge exchange,
tends to enhance the cascade rates by isotropizing the ISM turbulence
on a relatively long time scale.  This tends to modify the
characteristics of ISM turbulence which can be significantly different
from the Kolmogorov phenomenology of fully developed turbulence.  We
believe that, it is because of this enhanced nonlinear eddy
interaction time, that a large spectral transfer of turbulent energy
tends to smear the current sheets in the magnetic field fluctuations
and further cascade energy to the lower Fourier modes in the inertial
range turbulent spectra. Consequently it leads to a steeper power
spectrum. It is to be noted that the present model does not consider
an external driving mechanism, hence the turbulence is freely
decaying.  Driven turbulence, such as due to large scale external
forcing, e.g. supernova explosion, may force turbulence at larger
scales. This can modify the cascade dynamics in a manner usually
described by dual cascade process.

There are issues that need further exploration.  For example, neutrals
are observed to evolve isotropically in our coupled plasma-neutral ISM
turbulence model. A relevant question, whether their coupling to the
plasma fluid at large-scales reduces spectral anisotropy in the plasma
fluctuations, remains to be investigated. Furthermore, our model is
currently 2D. It is then intriguing whether variations in the third
dimension, i.e. 3D, introduce non trivial effects to our present 2D
studies because turbulence in 2D differs considerably from that in 3D,
particularly in the context of 3D neutral hydrodynamics fluid, where
vortex stretching effect (which is absent in 2D) tears apart the
large-scale structures and forbids the inverse cascade phenomena. By
virtue of latter, 2D and 3D neutral fluids possess distinct spectral
features characterized essentially by the number of the inviscid
quadratic invariants \citep{biskamp}. How coupling of the plasma and
neutral fluids in partially ionized ISM turbulence is modified by the
3D interactions remains a subject of our future investigations.

We add finally that our self-consistent model can be useful in
studying turbulent dynamics of partially ionized plasma in the
magnetosphere of Saturn and Jupiter where outgassing from moons and Io
and Encephalus introduces a neutral gas into the plasma.


\acknowledgments

 The support of NASA(NNG-05GH38) and NSF (ATM-0317509) grants is  acknowledged.






\clearpage

\begin{figure}
\plotone{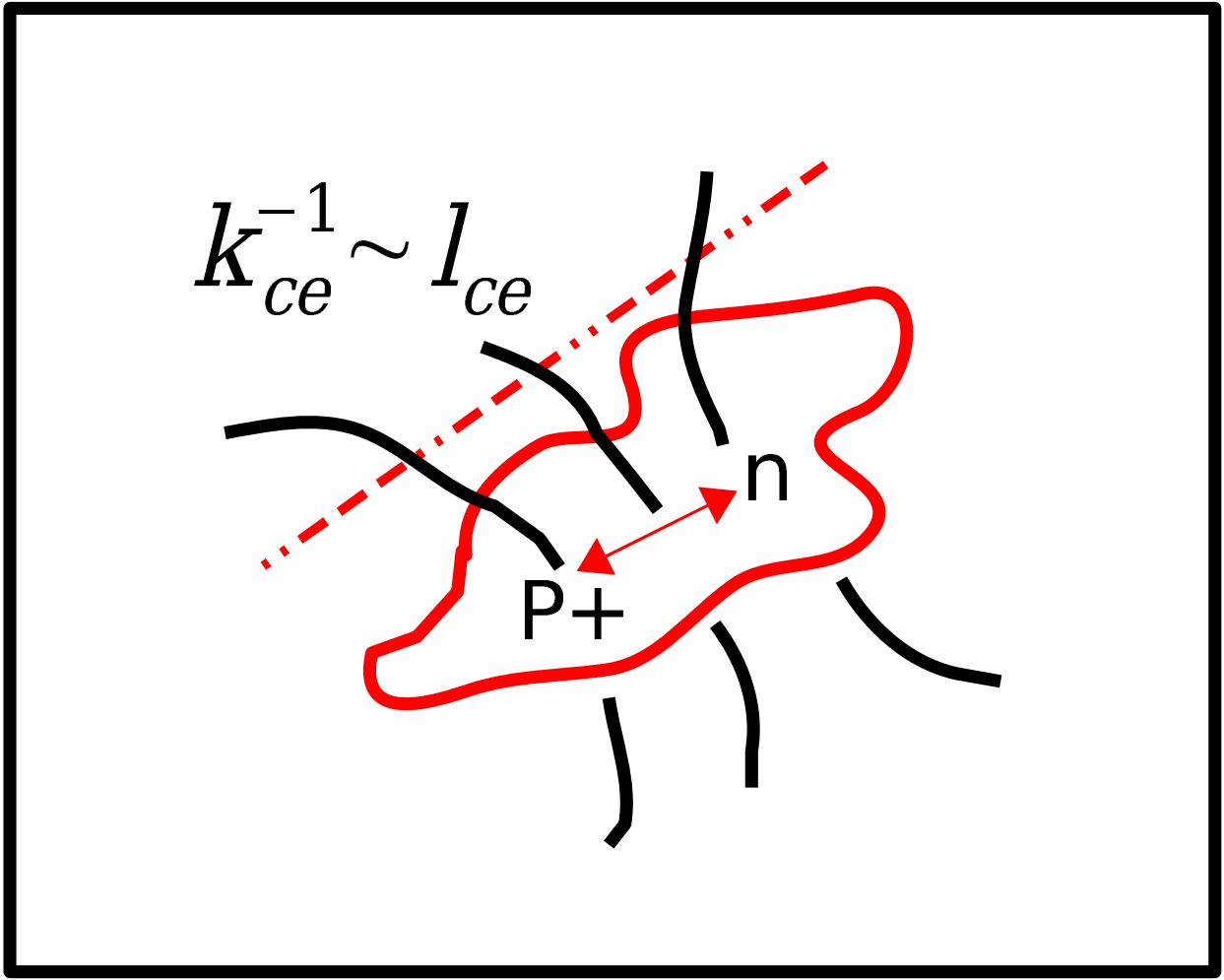} \figcaption{\label{fig1} Schematic
illustrating the significance of the charge exchange mode, which
corresponds to charge exchange interactions per unit area in the
computational domain. This mode, being different from the
characteristic turbulent mode $k$, is excited naturally when plasma
and neutral fluids are coupled in the ISM by charge exchange. }
\end{figure}

\begin{figure}
\plotone{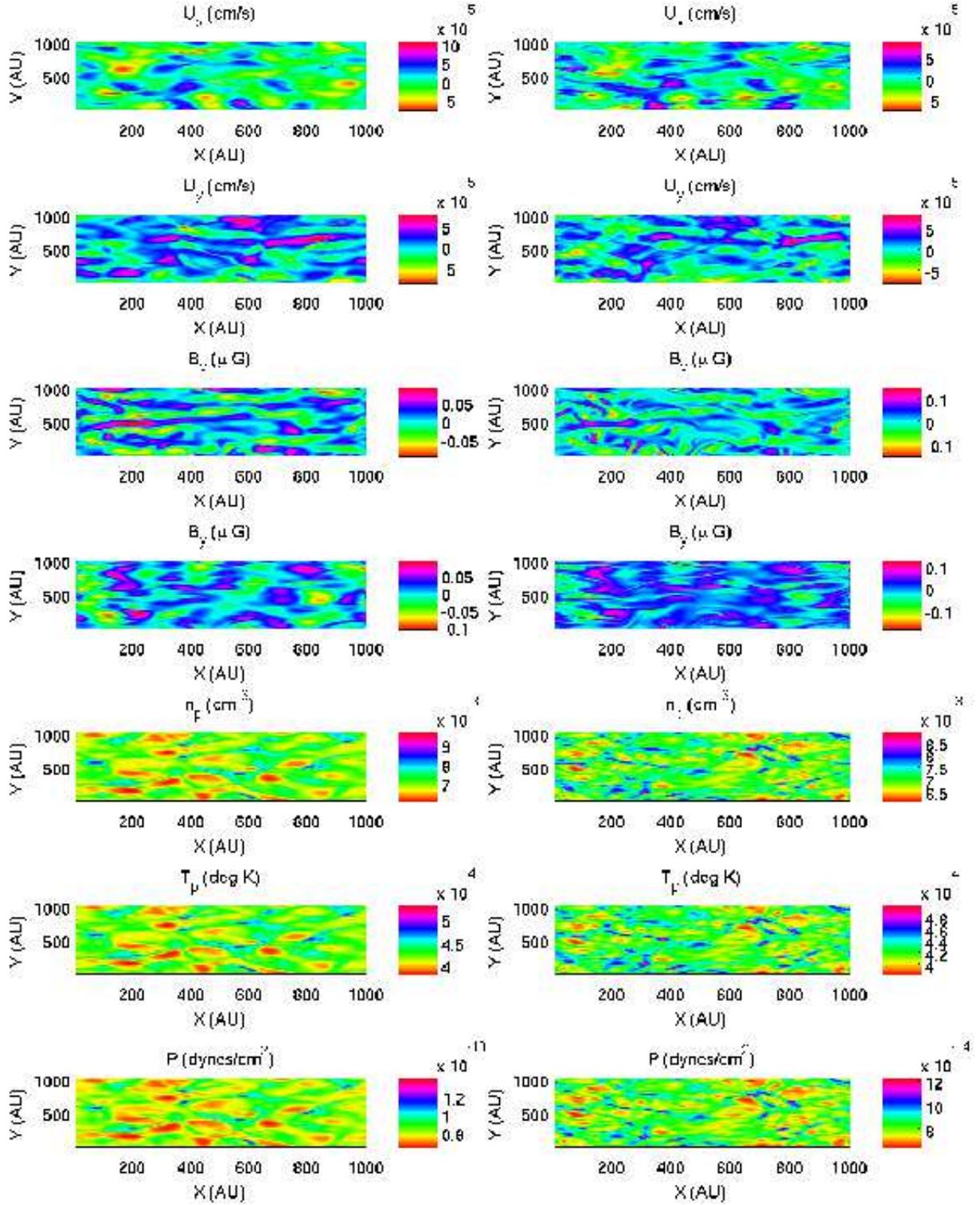} \figcaption{\label{fig3} ({\bf Left column}) Mode
  structures of plasma when coupled with the neutral ISM fluid.  The
  local ISM case is considered where typical LISM parameters are;
  $\gamma=5/3, n_{0_{plasma}} = 0.06 cm^{-3}, n_{0_{neutral}} = 0.18
  cm^{-3}, T = 6500 K, V_{th} = 10 km/s, \sigma = 6.9\times 10^{-15},
  l_0 = 200 AU, L({\rm box length}) = 5 l_0$. ({\bf Right column}) The
  corresponding quantities for an uncoupled ISM plasma simulation
  measured at a similar time step (i.e. $t=20$ in normalized units).
  Notice the magnetic field structures in the uncoupled case showing
  the formation of thin (or sheet-like) structures unlike the coupled
  system. Also notice the differences in the density fields.  }
\end{figure}

\clearpage

\begin{figure}
\plotone{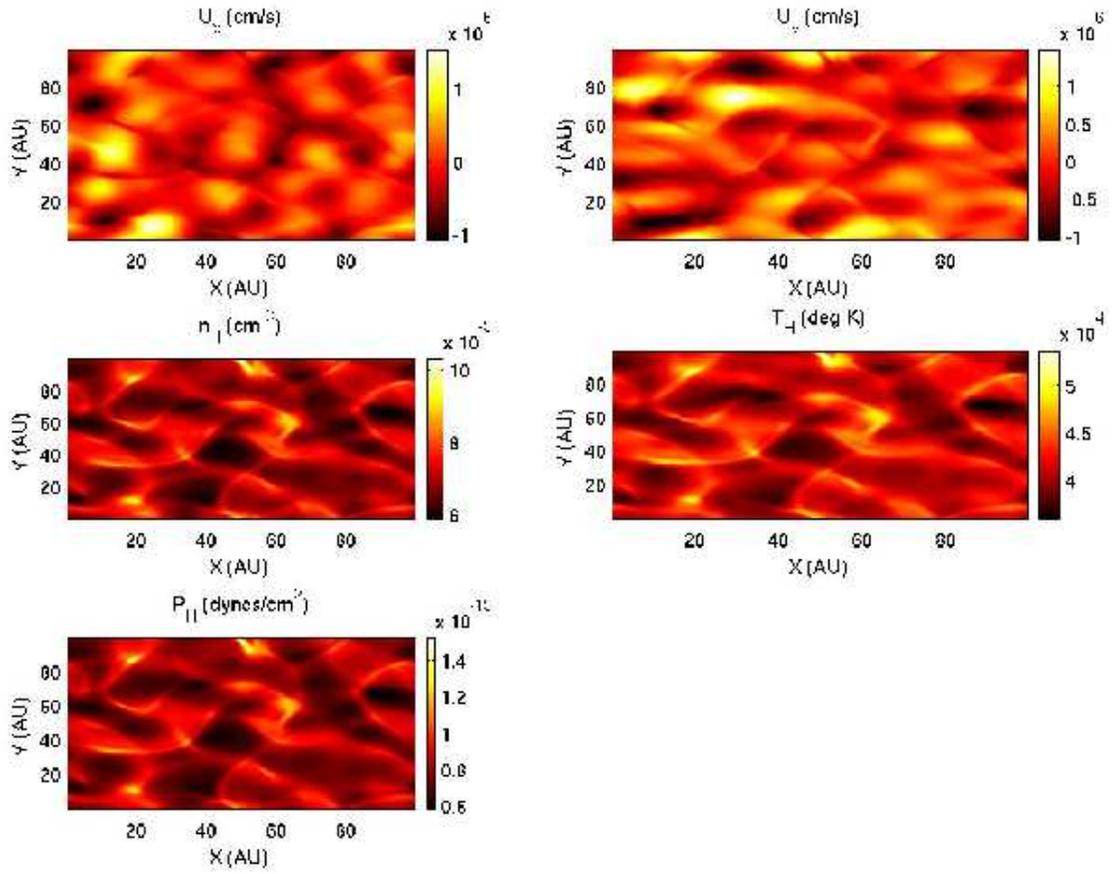} \figcaption{\label{fig4} The corresponding mode
  structures of neutral fluid in the presence of a plasma
  fluid. Simulation parameters are described in the previous figure.
}
\end{figure}

\clearpage

\begin{figure}
\plotone{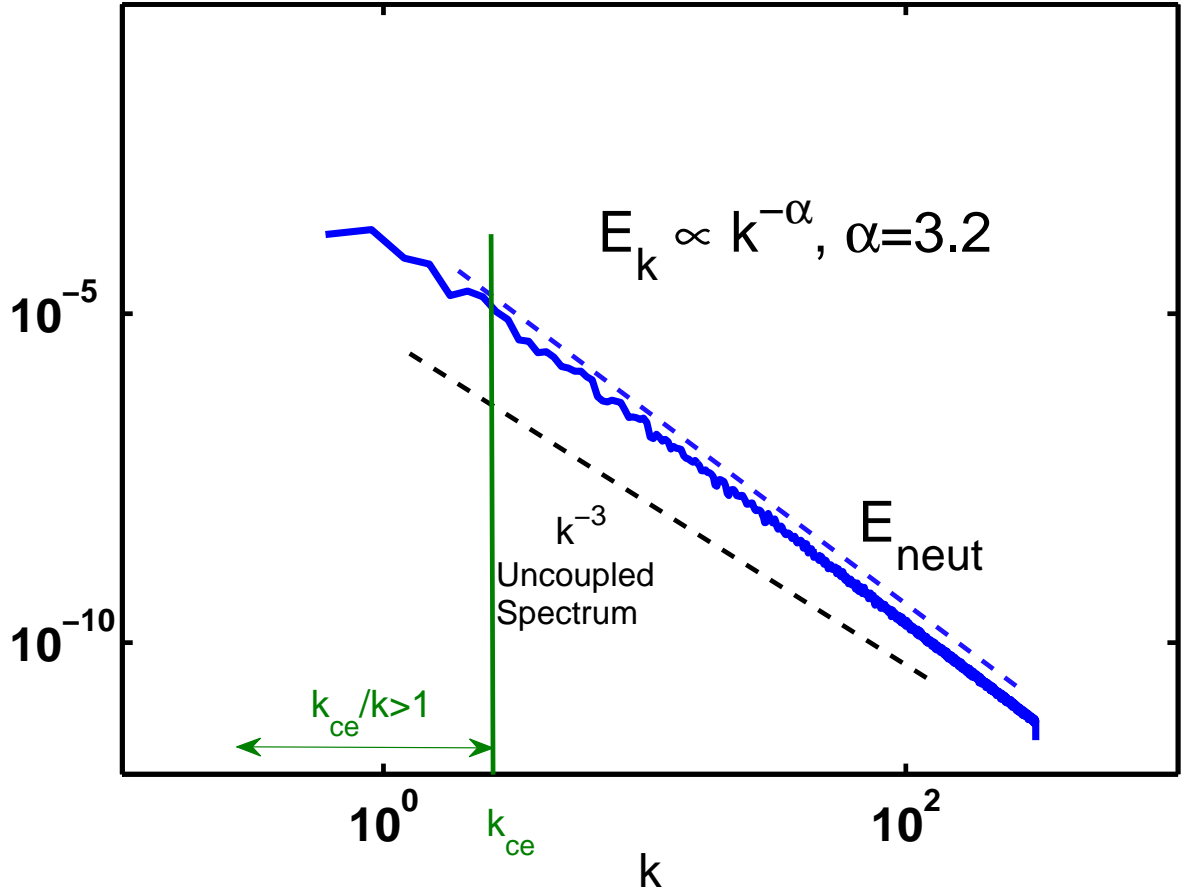} \figcaption{\label{fig5}The inertial range
  turbulent spectrum for a neutral fluid shows a $k^{-3.2}$ spectrum when
  coupled through charge exchange to a plasma fluid.  The spectral
  index exhibited by this spectrum is larger than that of an ordinary
  (hydrodynamic) fluid. The latter exhibits a Kolmogorov-like $k^{-3}$
  spectrum in the forward cascade regime (drawn schematically). The
  numerical resolution for our spectral studies is $1024^2$ modes in
  2D.}
\end{figure}

\clearpage

\begin{figure}
\plotone{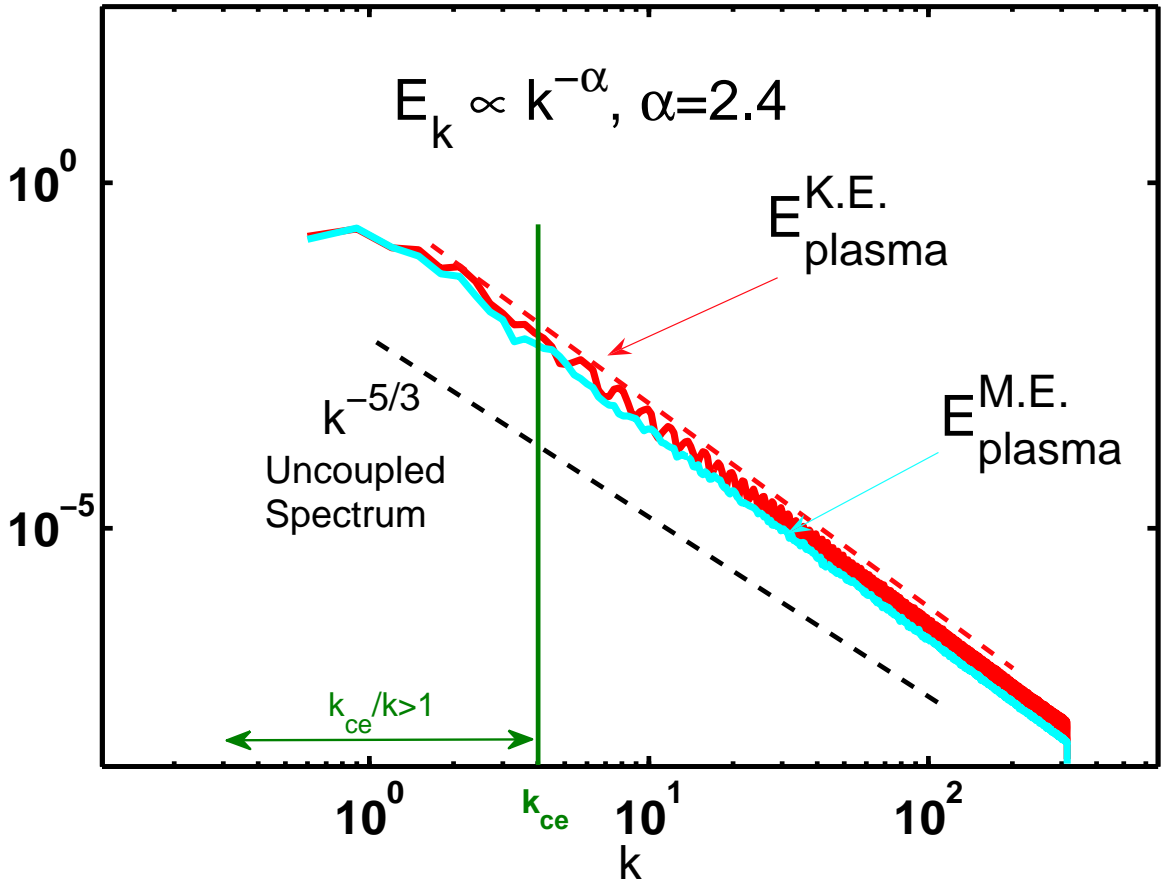} \figcaption{\label{fig6} A similar steepening of
  turbulent spectra occurs in the plasma fluid when coupled to the
  neutrals through charge exchange. An uncoupled magnetofluid plasma
  exhibits a Kolmogorov-like $k^{-5/3}$ spectrum in the forward
  cascade regime (drawn schematically).}
\end{figure}

\clearpage

\begin{figure}
\plotone{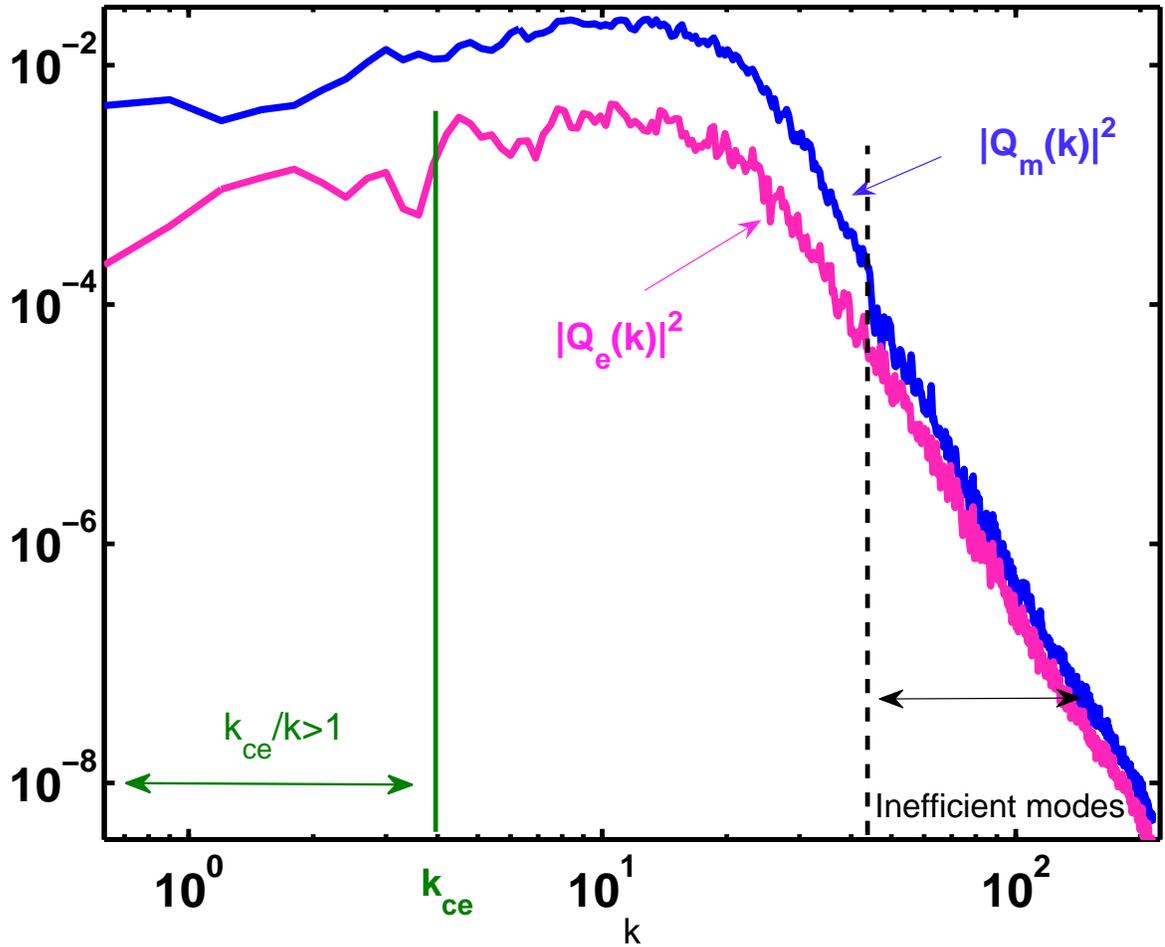} \figcaption{\label{fig7} 
Charge exchange spectrum associated with momentum and energy in both plasma
and neutral fluids. Clearly the energy cascades are dominated by large-scale
interactions, whereas small scales are not efficient enough to influence the
turbulent spectra in the local ISM.
}
\end{figure}

\clearpage

\end{document}